\documentclass[pra,reprint,superscriptaddress,nobibnotes,aps]{revtex4-1}
\usepackage{graphicx} 
\usepackage{appendix}
\usepackage{amsmath}
\usepackage{bm}
\usepackage{upgreek}
\usepackage{wasysym} 
\usepackage{hyperref}
\usepackage{siunitx}
\usepackage{svg}
\usepackage{tikz}
\usepackage{xcolor}
\usetikzlibrary{shapes.geometric}
\usetikzlibrary{decorations.pathreplacing}

\begin{document}

\title{Rapid all-optical loading of trapped ions using a miniaturised atom source}

\date{\today}
\author{L. Versini}
\thanks{These authors contributed equally to this work.}
\affiliation{Department of Physics, University of Oxford, Clarendon Laboratory, Parks Road, Oxford, OX1 3PU, United Kingdom}

\author{T. F. Wohlers-Reichel}
\thanks{These authors contributed equally to this work.}
\affiliation{Department of Physics, University of Oxford, Clarendon Laboratory, Parks Road, Oxford, OX1 3PU, United Kingdom}

\author{C. E. J. Challoner}
\affiliation{Department of Physics, University of Oxford, Clarendon Laboratory, Parks Road, Oxford, OX1 3PU, United Kingdom}

\author{T. Hinde}
\affiliation{Department of Physics, University of Oxford, Clarendon Laboratory, Parks Road, Oxford, OX1 3PU, United Kingdom}

\author{A. D. Rao}
\affiliation{Department of Physics, University of Oxford, Clarendon Laboratory, Parks Road, Oxford, OX1 3PU, United Kingdom}

\author{W. J. Hughes}
\affiliation{Optoelectronics Research Centre, University of Southampton, Southampton, S017 1BJ, UK}

\author{P. Drmota}
\affiliation{Department of Physics, University of Oxford, Clarendon Laboratory, Parks Road, Oxford, OX1 3PU, United Kingdom}
\affiliation{QFX Ltd., Centre for Innovation and Enterprise, Begbroke Science Park, Kidlington, OX5 1PF}

\author{T. H. Doherty}
\affiliation{Department of Physics, University of Oxford, Clarendon Laboratory, Parks Road, Oxford, OX1 3PU, United Kingdom}

\author{L. J. Stephenson}
\affiliation{QFX Ltd., Centre for Innovation and Enterprise, Begbroke Science Park, Kidlington, OX5 1PF}

\author{J. A. Blackmore}
\email[Corresponding e-mail:]{jacob.blackmore@physics.ox.ac.uk}
\affiliation{Department of Physics, University of Oxford, Clarendon Laboratory, Parks Road, Oxford, OX1 3PU, United Kingdom}

\author{J. F. Goodwin}
\affiliation{Department of Physics, University of Oxford, Clarendon Laboratory, Parks Road, Oxford, OX1 3PU, United Kingdom}
\affiliation{QFX Ltd., Centre for Innovation and Enterprise, Begbroke Science Park, Kidlington, OX5 1PF}

\begin{abstract}
    We characterise an efficient optically-heated neutral atom source for ion trapping. We observe loading rates of up to \SI{24(3)}{\per\second} with heating powers below \SI{85}{\milli\watt}, and demonstrate loading of a single ion in under \SI{30}{\second} with \SI{41.4(4)}{\milli\watt} of optical power in a room-temperature ion trap system with an ionisation probability of \num{1.50(5)e-5}.
    We calibrate a thermal model for the source's internal temperature by imaging the fluorescence of a collimated flux of neutral calcium that effuses from the source at various optical heating powers. 
    We show that the thermal performance of this source is mainly limited by radiative losses.
    We explore the effect of second-stage photo-ionisation laser power on the loading rate, and identify a path beyond the loading rates reported in this study. We predict that this source is also well-suited to a wide range of metals used in ion trapping.
\end{abstract}

\maketitle
\section{Introduction}
\begin{figure*}[t]
    \includegraphics[width=\textwidth]{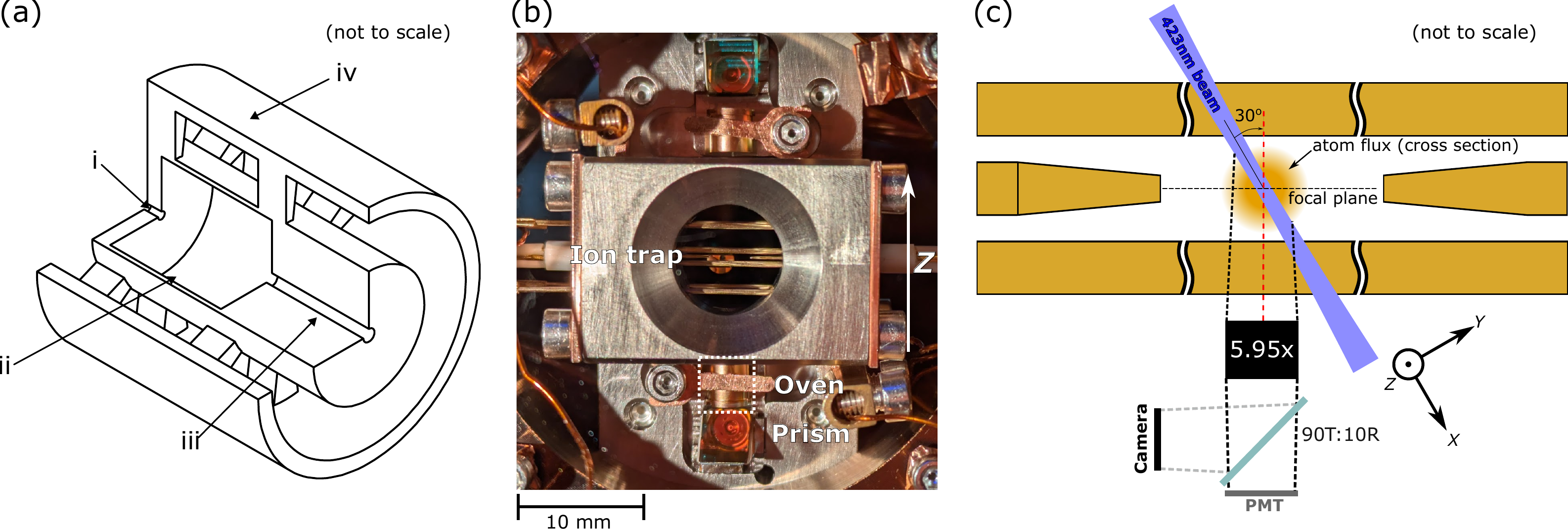}
    \caption{Simplified experimental setup (a) Construction of the oven. Shown schematically are: i, the aperture for input heating laser light; ii, the crucible for containing the atom source material; iii, the integrated collimator; and iv, the outer wall which is thermally isolated from the hot crucible, by a series of supports (shown here as radial spokes for simplicity). An outer jacket, used for mechanical stability, is omitted for clarity. (b) A photograph of the ion trap and oven mount installed in the chamber, the oven used in this work is highlighted. The electrodes are~\SI{0.5}{\milli\meter} in diameter and are spaced by~\SI{0.65}{\milli\meter} from the RF null. (c) An illustration depicting the midplane of the ion trap system. Simplified diagram of the imaging system; the central region is approximately to scale. Further, the coordinate axes used to analyse the atomic beam are rotated by \ang{30} to the trap axis and are consistent with (b).}
    \label{fig:neutral-setup}
\end{figure*}
Quantum technologies based on single trapped ions are of great interest at present, and they have been used successfully in the fields of metrology, sensing and quantum information~\cite{Harty2014, Clark2021, Hughes2025, Srinivas2021, Leu2023}, quantum simulation~\cite{Hempel2018, MacDonell2023, Navickas2025, Saner2025}, quantum-enhanced metrology~\cite{Valahu2025, Nichol2022} and in searches for new physics~\cite{Kozlov2018, Roussy2021}. As these technologies gradually move from the laboratory to deployable devices such as ion-based atomic clocks, quantum sensors and near-term quantum computers, it becomes increasingly important that ion trap systems can be operated reliably over time. 
 
To trap ions, neutral atoms have to be transported into the trapping volume, ionised, and cooled. The conventional methods of generating neutral atomic flux involve either resistively heated ovens~\cite{Lucas2004, Ballance2018}, ablation targets~\cite{White2022, Zimmermann2012, Hashimoto2006}, or magneto-optical traps (MOTs)~\cite{Sage2012}. We can compare these approaches by considering both the heat dissipated in the system and the ``loadable fraction'' of neutral atoms produced~\cite{smith2025}. Broadly, the loadable fraction depends on two parameters: the transverse spread of the atomic beam, which should be small compared to the trapping volume and the mean kinetic energy of the atomic beam, which should be low compared to the trap depth. 
The resulting loading rate additionally depends the number density, which should be as high as practical.

Resistively heated sources are widely used owing to their straightforward construction; however, their electrical connections impose unavoidable conductive losses, and conventional designs emit broad, diffuse beams of atoms. To compensate for these inefficiencies, they require significant powers, of order \SI{5}{\watt}, to heat the source metal to high temperatures (of order \SI{500}{\kelvin} for calcium)~\cite{Lucas2004, Ballance2018}. The electrical power dissipated in the vacuum chamber can cause temperature fluctuations, which are particularly deleterious for cryogenic experiments, and can disrupt high-precision experiments. There has been recent progress on reducing this effect using custom micro-fabricated resistively-heated thermal sources~\cite{Kumar2025}.

Ablation of a target by a high power laser has been demonstrated in a number of experiments~\cite{White2022,Zimmermann2012, Hashimoto2006}, and can enable the use of chemically-stable alkali-earth salt targets. Unfortunately, the resulting ablation plume contains a range of undesirable species, e.g., multiply-charged ions and molecules~\cite{Zimmermann2012}. Ablation of pure metallic sources can produce a cleaner plume, but the method still suffers from reliability issues owing to target degradation, and atoms are produced with a high average velocity resulting in a low loadable fraction~\cite{smith2025}.

Loading from a MOT offers key advantages in terms of the velocity distribution of the neutral atoms~\cite{Sage2012}; since the atoms are pre-cooled, the loadable fraction can approach unity. However, this method requires additional frequency-stabilised lasers, complex extensions to the apparatus, and restricts optical access to the ion trap.

In a previous work~\cite{Gao2021}, we demonstrated that laser-heated ovens provide an attractive alternative for neutral atomic flux generation. By removing the necessity of thermally conductive electrical connections onto the oven itself, the conductive losses to the surrounds were greatly reduced. Despite these improvements, this source was still limited by poor thermal efficiency necessitating in excess of \SI{300}{\milli\watt} to reach metal temperatures of \SI{550}{\kelvin}. Having been an uncollimated source, the loadable fraction of atoms produced was limited by the spatial extent of the beam. 

In this paper, we present a novel microfabricated optically heated atom source, from here on referred to as the ``oven'', which has been engineered to reduce its conductive and radiative losses allowing for high temperatures to be reached with a fraction of the power used in conventional resistively heated sources. The addition of an integrated high-aspect-ratio collimator ensures that the atomic flux is highly localised to the trapping region.

The paper is structured as follows: in~\autoref{sec:neutral} we describe the properties of the neutral atomic beam of calcium produced and model the thermal performance of the oven. In~\autoref{sec:ions} we demonstrate loading of single $^{40}$Ca$^+$ ions into a prototypical three-dimensional Paul trap and quantify the loading rate. Finally, in~\autoref{sec:conclusions} we provide an outlook on possible operational regimes accessible with this oven. We also discuss applications to other atomic species and the technical challenges they face.

\section{Neutral atomic beam characterisation}\label{sec:neutral}

We characterise the atomic number density by operating the oven at sufficient fluxes where resonance fluorescence can be detected from neutral calcium atoms with a conventional imaging system and photon counting electronics. From this, we calculate the atomic flux as a function of heating power, which we then use to infer the temperature inside the oven.

\subsection{Experimental setup}
The oven is constructed from UV-fused silica parts machined using selective laser-induced etching~\cite{Bellouard2012}.
We illustrate the key features of the oven schematically in~\hyperref[fig:neutral-setup]{\autoref*{fig:neutral-setup}(a)}. Heating light is supplied through a narrow aperture on the rear face of the oven (i), which illuminates the calcium metal contained within a central crucible (ii). The heated metal produces vapour, which effuses through a high-aspect-ratio collimator (iii). To reduce conduction to the surrounding apparatus the outer wall of the oven (iv) is thermally isolated from the crucible by a numerous spoke-like supports, which take a meandering path from the crucible to the outer wall. The spokes have cross-sectional areas ranging from \SIrange{0.0025}{0.015}{\milli\meter^2} and crucible-to-wall path lengths approaching \SI{10}{\milli\meter}~\footnote{note that the supports have been greatly simplified in~\hyperref[fig:neutral-setup]{\autoref*{fig:neutral-setup}(a)} to improve clarity}.
To reduce the thermal emissivity of the oven, each of these components is coated with a Ti/Au stack in a magnetron sputterer, where the coating thickness is controlled to ensure thermal conductance remains dominated by the fused silica bulk. We mount the oven in a titanium V-groove, mechanically retained by an oxygen-free copper clamp, approximately \SI{10}{\milli\meter} from the trap centre (see~\hyperref[fig:neutral-setup]{\autoref*{fig:neutral-setup}b}).

The measurements of the atomic flux are performed in an ultra-high vacuum chamber into which a linear Paul (RF) trap, described in detail in~\autoref{sec:ions}, is mounted. The atoms emerge from the oven collimator and propagate along the vertical axis of the chamber, which we label $Z$. We use excitation laser light resonant with the $4\mathrm{s}^{2} \, {}^{1}\mathrm{S}_{0}$ $\leftrightarrow 4 \mathrm{s} 4\mathrm{p} \, {}^{1}\mathrm{P}_{1}$ transition at \SI{423}{\nano\meter} in \textsuperscript{40}Ca, to probe the atomic beam. The propagation axis of this laser beam defines the $X$ axis. We image the atoms along the quantisation axis, given by an \SI{0.8}{\milli\tesla} magnetic field, which lies at \SI{30}{\degree} to $X$. The path of the excitation laser through the trapping region is shown schematically in~\hyperref[fig:neutral-setup]{\autoref*{fig:neutral-setup}c}. The excitation laser has beam waist \SI{31(2)}{\micro\meter} and power \SI{1.6(2)}{\micro\watt} or \SI{5.2(5)}{\micro\watt} depending on the dataset (shown in~\hyperref[fig:neutral-density]{\autoref*{fig:neutral-density}}).
The laser light at \SI{785}{\nano\meter} used to heat the oven is directed onto the rear side of the oven via an in-vacuum dielectric prism (see Fig.~1b).
\begin{figure}[t]
\includegraphics[width=\linewidth]{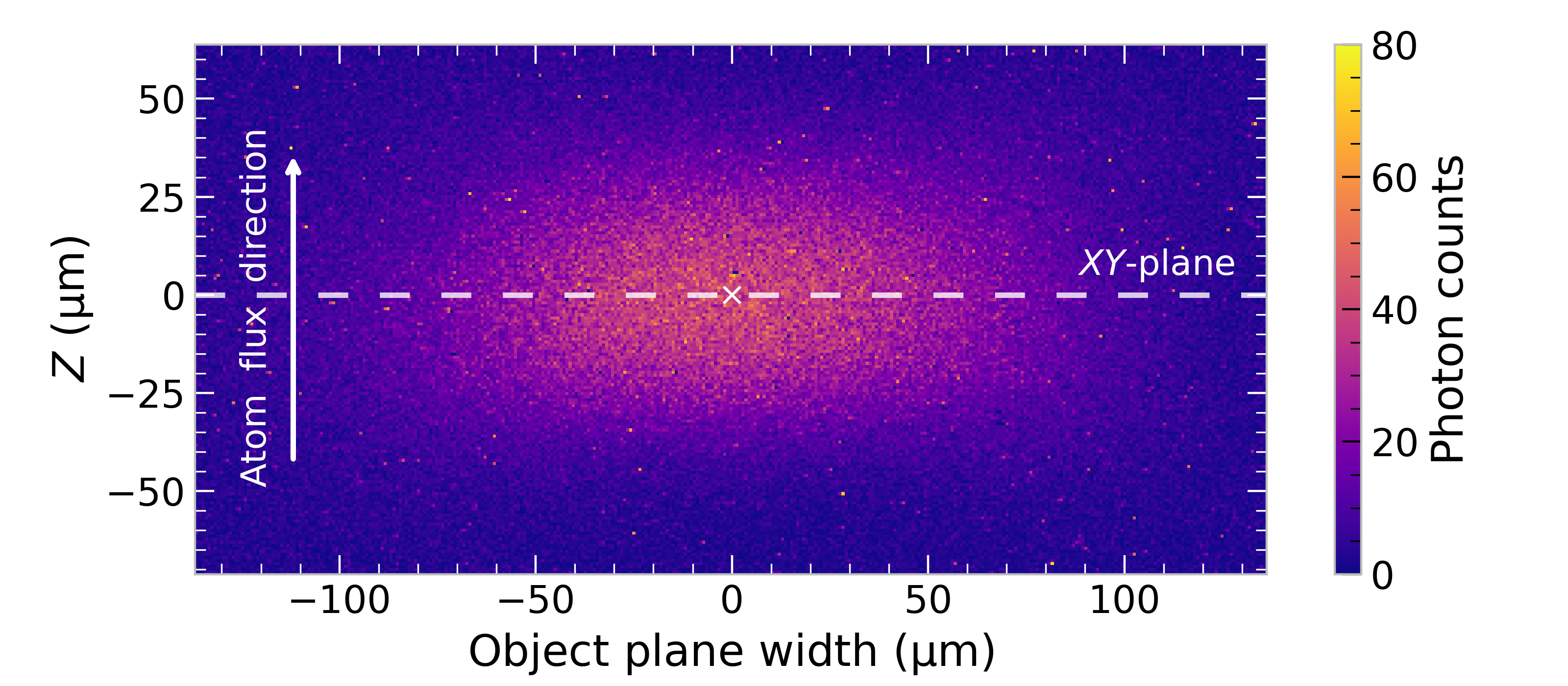}
    \caption{A false-colour image of resonant fluorescence from neutral $^{40}$Ca taken using the imaging system shown in~\hyperref[fig:neutral-setup]{\autoref*{fig:neutral-setup}c}. The axes are scaled to the object plane by the known magnification of $M=5.95(2)$. The white arrow is aligned with the z axis in \autoref{fig:neutral-setup} and labels the atom flux direction. The dashed yellow line shows the projection of the $XY$-plane labelled in \autoref{fig:neutral-setup} onto the image plane. The wave vector of the excitation laser beam is in $XY$ and makes an angle of $30^\circ$ with the optical axis. The white cross $\times$ shows the center of the fitted distribution.}
    \label{fig:neutral-image}
\end{figure}
\begin{figure}[!t]
    \centering
    \includegraphics[width=0.935\linewidth]{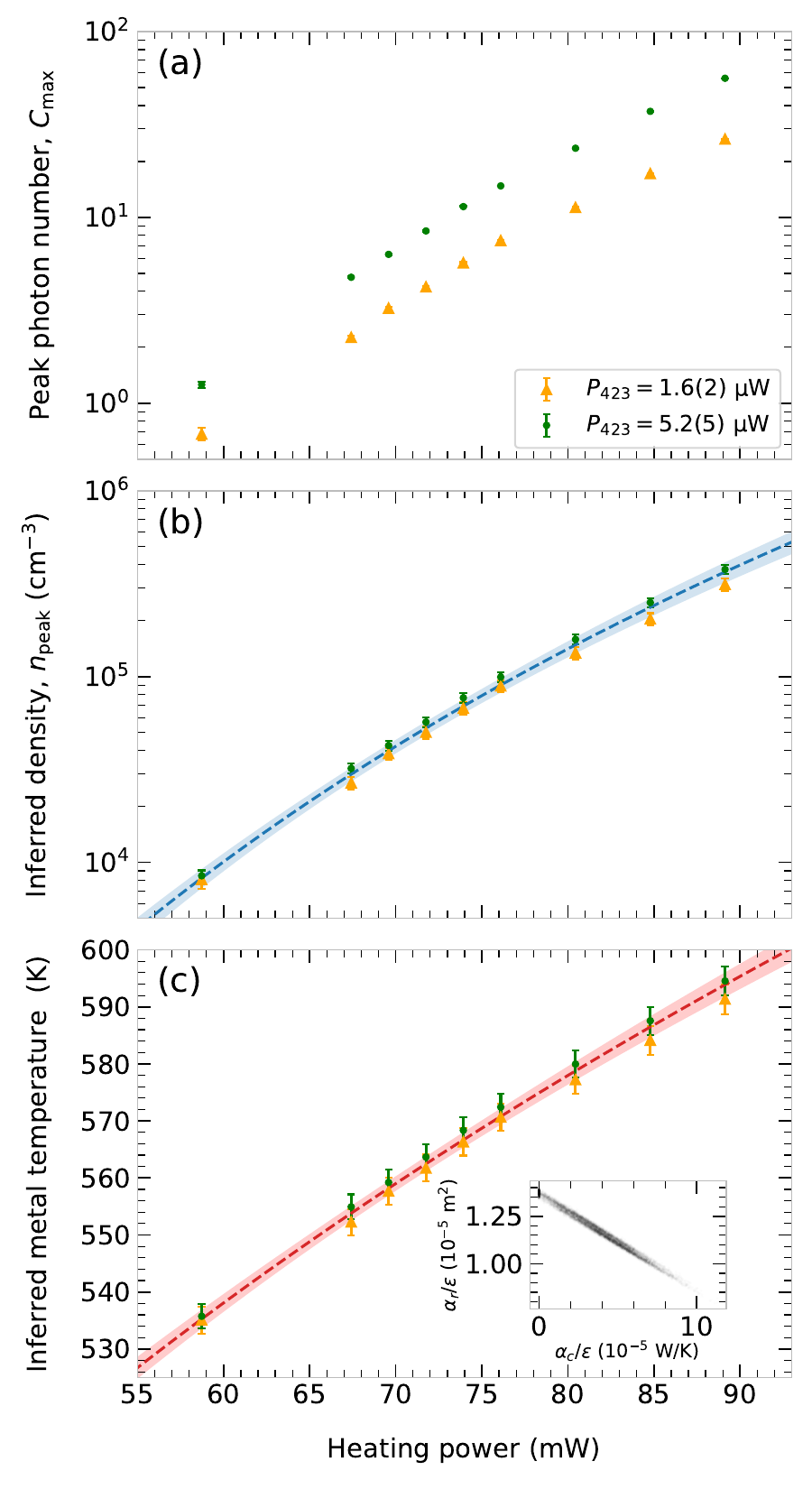}
    \caption{Determination of the oven's thermal performance from fluorescence measurements. 
    (a) Peak photon count $C_\mathrm{max}$ extracted from Gaussian fits of fluorescence images as shown in \autoref{fig:neutral-image}. Datasets are shown for two excitation laser powers, \SI{1.6(2)}{\micro\watt} (yellow triangles) and \SI{5.2(5)}{\micro\watt} (green circles). (b) The peak atomic density is inferred from the camera data in (a). The dashed blue line shows $n_{\mathrm{peak}}$ as inferred by the thermal model. (c) Synchronous camera and PMT readings are combined to quantify the total atomic flux and deduce the crucible temperature. The red line is the best fit of the thermal model to the temperature data.
    The inset shows the distribution of the model parameters obtained by bootstrapping, each point corresponds to a unique set of parameters. The conductive loss coefficient $\alpha_\mathrm{c}/\varepsilon = \SI{4(3)e-5}{\watt\per\kelvin}$ could not be well established due to the thermal response being dominated by radiative losses in this regime. The radiative loss coefficient is fitted to be $ \alpha_{\mathrm{r}}/\varepsilon = \SI{1.2(1)e-5}{\meter\squared}$. The thermal model obtained from data shown in (c) is used to predict $n_\mathrm{peak}$ in (b). Shaded areas represent $1\sigma$ confidence intervals.}
    \label{fig:neutral-density}
\end{figure}
\subsection{Imaging system}
The fluorescence of the neutral atoms is collected by an infinity corrected objective (Mitutoyo ${\mathrm{MY10X-803}}$). A telescope is used to magnify the image to $M=5.95(2)$, with the light split between a photo-multiplier tube (PMT) (H10682-210-MOD, Hamamatsu) and CMOS camera (ORCA Quest 1, Hamamatsu). Background scatter is rejected by an adjustable aperture placed at an intermediate focus of the telescope. The total detection efficiency of the system, including the quantum efficiency of the PMT, at \SI{397}{\nano\meter} and \SI{423}{\nano\meter} was measured to be \SI{0.25(1)}{\percent} and \SI{0.42(3)}{\percent}, respectively.

\subsection{Number density and temperature}
To quantify the thermal performance of the oven we determine the total flux of neutral atoms that pass through the trapping region. We resonantly excite the $^1S_0 \leftrightarrow {^1P_1}$ transition and count the number of photons incident on the camera within a collection duration $t_\mathrm{c} = \SI{120}{\second}$. 
The peak value of the photon distribution $C_\mathrm{max}$ imaged on the camera emerges from the intersection of the excitation laser and the atomic beam as depicted in~\autoref{fig:neutral-image}. $C_\mathrm{max}$ is determined by fitting a two-dimensional Gaussian to the image.
In these images, the vertical ($Z$) extent of the detected fluorescence is limited by the excitation laser power and $1/e^2$ radius. In the horizontal direction, a projection of the $XY$-plane, the dependence is much more complex owing partly to the finite depth-of-field of the imaging system, the extent of the atomic beam, and the 30$^\circ$ angle between the excitation laser and the optical axis.

\hyperref[fig:neutral-density]{\autoref*{fig:neutral-density}a} shows the measured peak photon number $C_\mathrm{max}$ as a function of the heating laser power for two different saturation parameters $s=I/I_{\mathrm{sat}}$ corresponding to a peak intensity of the excitation laser beam of $I\approx3 I_\mathrm{sat}$, and $I\approx 10 I_\mathrm{sat}$.
The saturation intensity $I_{\mathrm{sat}}=A_{\mathrm{SP}}\hbar \omega_0^3/(12\pi c^2) \approx \SI{600}{\watt\per\meter\squared}$ of the $^1S_0 \leftrightarrow {^1P_1}$ transition is defined by the corresponding Einstein coefficient $A_{\mathrm{SP}}$ and transition frequency $\omega_0$\cite{lurio_lifetime_1964}.

To convert these measurements to the absolute atomic density, we establish a model for the fluorescence rate in the interaction region. The number of photons scattered per second in a region of volume $\mathrm{d}^3\vec{r}$ summed over the states in the $^1P_1$ manifold is given by
\begin{equation}\label{eq:infinitesimal-scatter}
   \gamma(\vec r)\ \mathrm{d}^3 \vec{r} = \ n(\vec{r})A_{\mathrm{SP}} {P_{\mathrm{T}}}\left(s(\vec{r})\right) \mathrm{d}^3 \vec{r} \ ,
\end{equation}
where $\gamma(\vec r)$ is the scattering rate density and $n(\vec r)$ is the number density of $^{40}\mathrm{Ca}$, where we have assumed that the contribution from other isotopes is negligible due to low-power off-resonant driving. The steady-state population $P_{\mathrm{T}}$ of the excited state can be found by solving the master equation of an atom interacting with the excitation laser. The function $P_{\mathrm{T}}$ is adjusted for the Doppler broadening caused by a misalignment of $\sim\SI{1}{\degree}$ between the excitation laser and the atomic beam (for a full derivation see ~\hyperref[app:steady-state]{Appendix~\ref*{app:steady-state}}). 
By integrating over the optical axis of the imaging system (shown by the red line in~\hyperref[fig:neutral-setup]{\autoref*{fig:neutral-setup}c}) we relate $C_\mathrm{max}$ to the scattering rate density through
\begin{multline}\label{eq:c_max}
 C_{\mathrm{max}} =\eta^{(\mathrm{cam})}_{\mathrm{ce}}t_{\mathrm{c}}\frac{A_{\mathrm{px}}}{M^2} \int_{-\infty}^{+\infty} \sqrt{1+m^2}\ \gamma(\vec r_\xi) \ \mathrm{d}\xi \ ,
\end{multline}
where $\eta^{(\mathrm{cam})}_{\mathrm{ce}}$ is the collection efficiency of the camera, the ratio of pixel area over the square of the magnification $A_{\mathrm{px}}/M^2=\left[\SI{4.6}{\micro\meter}/5.95(2)\right]^2$ yields the area of the pixel in the object plane of the imaging system, and $\vec r_\xi=(\xi, m\xi, 0)$. In the above, $\sqrt{1+m^{2}}\ d\xi$ is the length of the infinitesimal line element, where $m=\tan{(\SI{30}{\degree})}$ arises from the angle between the excitation laser and the imaging system. We approximate the density of atoms to be uniform over the narrow laser waist; we estimate that the impact of this simplification on $n(\vec{r})$ is less than the contributions from other experimental error sources. Using this, we rearrange~\autoref{eq:c_max} and~\autoref{eq:infinitesimal-scatter} to obtain an expression for peak density $n_{\mathrm{peak}}$ as a function of $C_{\mathrm{max}}$,
\begin{equation}
    n_{\mathrm{peak}} = \frac{M^2 C_{\mathrm{max}}}{\eta_{\mathrm{ce}}^{(\mathrm{c})}t_\mathrm{c} A_{\mathrm{px}} A_{\mathrm{SP}} \int_{-\infty}^{+\infty} \sqrt{1+m^2}\ {P_{\mathrm{T}}} \mathrm{d}\xi} \ .
\end{equation}
The inferred values of the peak atomic density are reported in \hyperref[fig:neutral-density]{\autoref*{fig:neutral-density}b}. 

Determining the temperature inside the crucible requires knowledge of the \emph{total} atomic flux leaving the oven, whilst our previous method yielded the \emph{peak} atomic density. Due to the \SI{30}{\degree} angle between the imaging system and the excitation laser and the finite depth of focus, the extremes of the interaction region are no longer sharply imaged. This means that the spatial information in the image cannot be easily converted to an atomic flux. However, by using the PMT in conjunction with the camera we can measure the integrated photon counts across the whole region. Then, we fit a model of the atomic density $n(\vec r)$ to the measurements shown in \hyperref[fig:neutral-density]{\autoref*{fig:neutral-density}b} to determine the total atomic flux. We consider $n(\vec{r})$ to be uniform in $Z$ over the diameter $2 \times w_0= \SI{62(4)}{\micro\meter}$ of the excitation laser and to be Gaussian-like in the plane perpendicular to $Z$:
\begin{equation}
    n(\vec r; w_{\mathrm{a}}) = n_{\mathrm{peak}}\ e^{-\frac{2(x^2 + y^2)}{w_{\mathrm{a}}^2}} \ .
\end{equation}
In the above, $w_{\mathrm{a}}$ is the width of the atomic beam in the object plane and is defined as the $1/e^2$ radius of the distribution, following the same convention as Gaussian laser beams. The total counts registered by the PMT are given by the integral of \autoref{eq:infinitesimal-scatter} over all space,
\begin{equation}\label{eq:CPMT}
    C_{\mathrm{PMT}}(w_{\mathrm{a}}) = \eta_{\mathrm{ce}}^{\mathrm{(PMT)}} n_{\mathrm{peak}} A_{\mathrm{SP}} \int \mathrm{d}^3\vec r e^{-\frac{2(x^2 + y^2)}{ w_{\mathrm{a}}^2}} P_{\mathrm{T}}(s(\vec r)) \ .
\end{equation}
We fit~\autoref{eq:CPMT} to the densities reported in ~\hyperref[fig:neutral-density]{\autoref*{fig:neutral-density}b} and the accompanying PMT measurement, yielding a best fit value of $w_{\mathrm{a}} = \SI{218(15)}{\micro\meter}$. 

The flux produced by the source is well approximated by the integral of the density over the $(x,y)$ plane multiplied by the average velocity of the atoms and corrected for isotopic abundance, $a_{40}=96.9\%$ \cite{berglund_isotopic_2011}, i.e.,
\begin{equation}
    \Phi_{\mathrm{out}} = \frac{1}{2}\pi  w_{\mathrm{a}}^2 n_{\mathrm{peak}} \bar{v} /a_{40}\ .
\end{equation}
By relating the output flux to the input flux through the transmittance $p_t$ of the collimator, as shown in~\hyperref[app:flux]{Appendix~\ref*{app:flux}}, we can relate the temperature inside the crucible to the measured peak number density:
\begin{equation}\label{eq:vapour}
    \frac{P(T)}{k_B T}=\frac{3\pi^2}{4p_t}\frac{w_{\mathrm{a}}^2}{A_{\mathrm{coll}}} \frac{n_{\mathrm{peak}}}{a_{40}} \ ,
\end{equation}
where $P(T)$ is the calcium vapour pressure in the crucible at temperature $T$ and $A_{\mathrm{coll}}=\pi(\SI{25}{\micro\meter})^2$  is cross section area of the collimator. For each point in \hyperref[fig:neutral-density]{\autoref*{fig:neutral-density}b},~\autoref{eq:vapour} is numerically solved for $T$.
The result of this analysis is shown in \hyperref[fig:neutral-density]{\autoref*{fig:neutral-density}c}.

To predict the peak densities beyond the regime we were able to probe, we construct a simple model of the thermal behaviour of the oven. This model accounts for the conductive and radiative losses which must be balanced by the fraction $\varepsilon$ of heating power $P_{\mathrm{in}}$, which is coupled into the crucible, to maintain thermal equilibrium:
\begin{equation}\label{eq:thermal-model}
    P_{\mathrm{in}}   = \frac{\alpha_\mathrm{c}}{\varepsilon}(T-T_{\mathrm{env}}) + \frac{\alpha_{\mathrm{r}}\sigma_\mathrm{B}}{\varepsilon} (T^4 - T^4_{\mathrm{env}}) \ ,
\end{equation}
In the above, $T_{\mathrm{env}}=295\ \mathrm{K}$ is the environment temperature and $\sigma_B$ is the Stephan-Boltzmann constant. The fitting of $\alpha_\mathrm{c}/\varepsilon$ and $\alpha_\mathrm{r}/\varepsilon$ is shown as a probability distribution in the inset in \hyperref[fig:neutral-density]{\autoref*{fig:neutral-density}c}. Note that we do not separate contributions from $\varepsilon$ and the thermal losses. For the heating power regime we were able to probe, we cannot uniquely identify the parameters individually as they are strongly correlated. To estimate the uncertainty on our fit we use a Monte Carlo bootstrapping method which resamples our experimental data, yielding $\alpha_\mathrm{c}/\varepsilon= \SI{4(3)e-5}{\watt\per\kelvin}$ and $\alpha_\mathrm{r}/\varepsilon=\SI{1.2(1)e-5}{\square\meter}$.

Despite the limitation on our fit, the parameters are sufficiently constrained to show that the thermal performance is limited by radiative losses. 
Furthermore, we have observed that a significant density of neutral atoms can be produced with modest heating laser power, and our simple model reveals that this can only be reasonably improved by reducing the radiative losses further through an improved coating process.

\section{Ion loading}\label{sec:ions}

The oven is designed to facilitate rapid and efficient loading of ions into a trap. As the atoms effusing from the source have temperatures below \SI{600}{\kelvin}, we expect that almost all ionised atoms will be trapped in traps with depths of order \SI{1}{\electronvolt}.

At low heating powers, the neutral atom flux through the trapping region becomes too low to be detectable as the average number of atoms in the overlap volume is less than one. The flux is however still sufficient to load individual ions at a reasonable rate. As we have a thermal model for source performance, by collecting loading rate data over a similar range of optical heating powers, we can determine the relationship between number density of neutral atoms and ion loading rate. This then allows the ion loading rate to be used to validate the thermal model at lower heating powers where a direct measurement of the number density was not possible.

\subsection{Experimental setup}
This work utilises a linear RF Paul trap, constructed from \SI{0.5}{\milli\meter} diameter rods and \SI{0.2}{\milli\meter} diameter pointed endcaps separated by \SI{3.9}{\milli\meter}. An image of the trap is shown in~\hyperref[fig:neutral-setup]{\autoref*{fig:neutral-setup}b} and a schematic is shown in~\hyperref[fig:neutral-setup]{\autoref*{fig:neutral-setup}c}. It is operated using a monopolar RF drive at \SI{14.69}{\mega\hertz} with peak-to-peak voltage of approximately \SI{400}{\volt}. The shortest ion-electrode distance is \SI{0.65}{\milli\meter}. This combination of geometry and voltages results in typical axial and radial trap frequencies of \SI{200}{\kilo\hertz} and \SI{1}{\mega\hertz}, respectively. An additional four electrodes, with an ion-electrode distance of \SI{4}{\milli\meter}, are used with voltages up to \SI{10}{\volt} to compensate for stray electric fields. These compensation voltages were set using a photon-correlation technique that allows for micromotion minimisation~\cite{Berkeland1998}.

The laser beams form an angle of \SI{30}{\degree} with respect to the optical axis. The $1/e^{2}$ radii for the two beams at the trap are \SI{29(2)}{\micro\meter} and \SI{41(2)}{\micro\meter} for the 397-nm cooling and 866-nm repump laser, respectively. The second step ionisation laser, which has a nominal wavelength of \SI{375}{\nano\meter}, co-propagates with the 423-nm excitation laser and it has a beam waist of \SI{184(2)}{\micro\meter}, resulting in an approximately uniform UV field over the interaction region.

\begin{figure}[t]
    \includegraphics[width=1\linewidth]{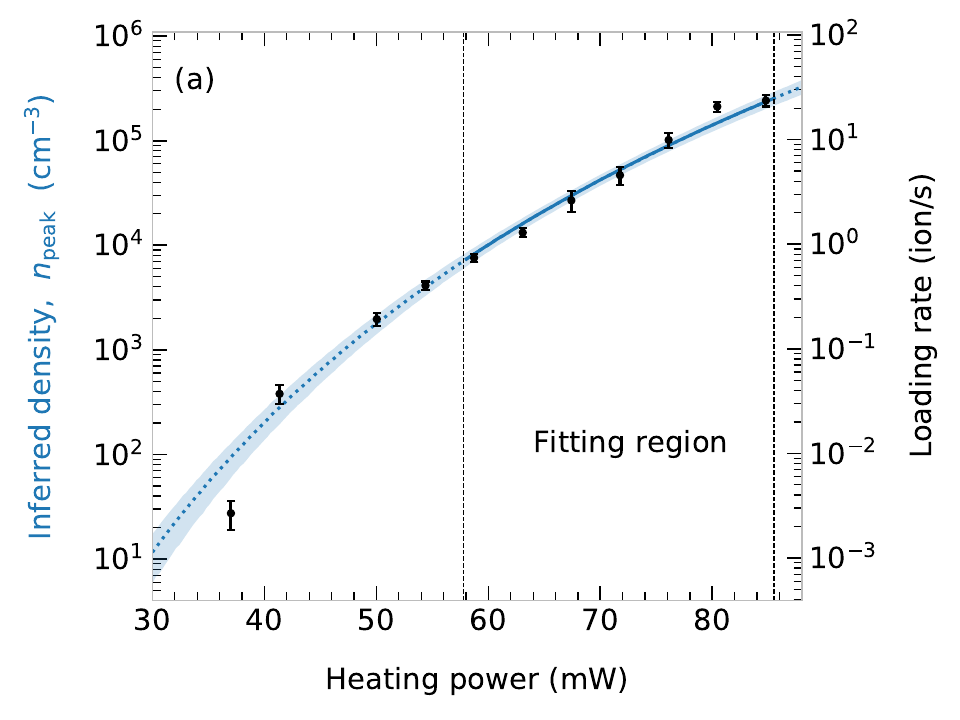}
    \includegraphics[width=0.90\linewidth]{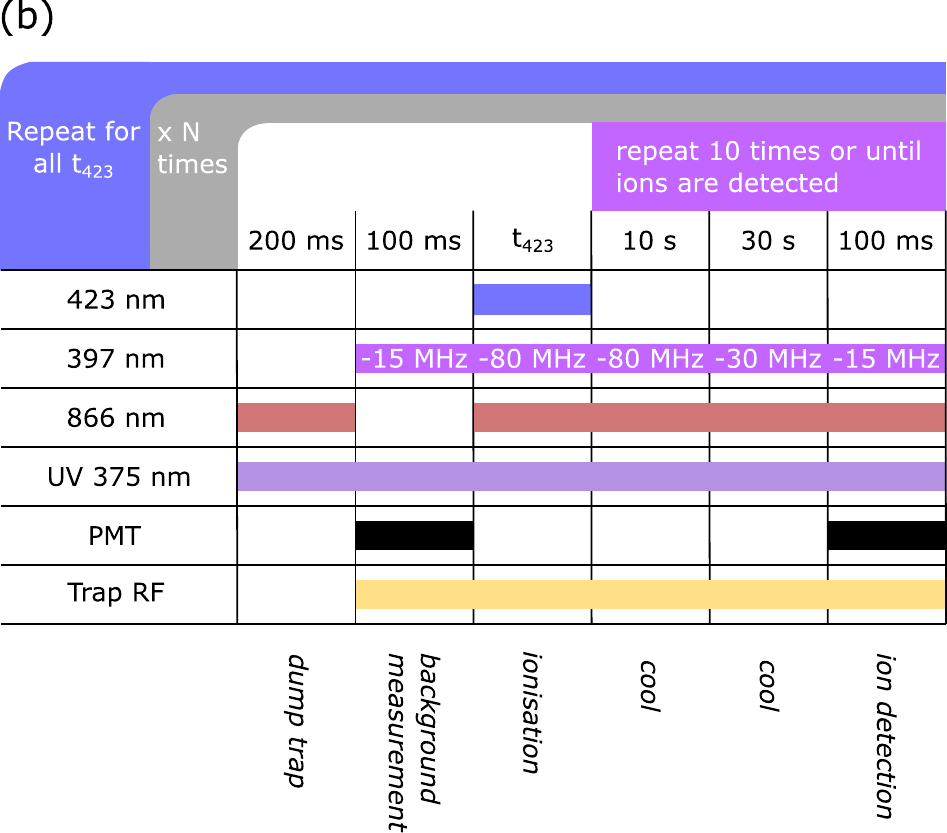}
    \caption{Loading rate comparison to number density and pulse scheme. (a) Loading rate of $^{40}\mathrm{Ca}^+$ ions (black circles, right axis) versus varying heating power with modelled central number density (dashed line, left axis). Dashed vertical lines indicate the region used to determine the relationship between the loading rates and number densities. The thermal model is reproduced from~\autoref{fig:neutral-density}. The shaded area represents one standard deviation for the thermal model. The ion loading rate data follows the same trend as the modelled number density. (b) Pulse sequence for the loading rate measurement. The repump (\SI{866}{\nano\meter} wavelength) laser is briefly turned off at the beginning to measure background counts. The excitation (\SI{423}{\nano\meter} wavelength) laser pulse is scanned in duration followed by a cooling and probe cycle using  \SI{80}{\mega\hertz},  \SI{30}{\mega\hertz} and \SI{15}{\mega\hertz} red detuned pulses of cooling (\SI{397}{\nano\meter} wavelength) laser light. Detection is carried out during the \SI{15}{\mega\hertz} red detuned cooling laser pulse in order to check if ions have been loaded. }
    \label{fig:loading_rate_number_density}
\end{figure}

\subsection{Loading rate characterisation}

Loading rate measurements are performed by scanning the pulse duration of the excitation laser beam. Before the experiment begins, the trapping potential is briefly switched off to allow any ions to escape from the trapping region. During the entirety of one experimental repetition, the cooling, repump and ionisation lasers are left on. By pulsing the excitation laser, the ionisation dynamics can be separated from the cooling dynamics in the trap. During all loading rate measurements, the oven was heated with the specified optical power and allowed to reach thermal equilibrium.
Following the ionisation step, in order to cool the ions into the trap, cooling cycles consisting of two sequential pulses of cooling laser light at different detunings are used, first at 80 MHz and then at 30 MHz, both red detuned from the $S_{1/2} \leftrightarrow  P_{1/2}$ transition. 
After each cooling cycle, the fluorescence is measured on the PMT with the cooling laser 15 MHz red detuned from the $S_{1/2} \leftrightarrow  P_{1/2}$ transition. If the counts are sufficiently above the background, this indicates the presence of at least one ion. After the cooling cycle finishes, the number of ions trapped is counted and the next cycle begins.
We determine the total number of ions by counting the number of photons on a PMT within an exposure time of \SI{100}{\milli\second} and dividing this by the signal we measure from one ion. 
As no more than six ions were loaded in any given repetition, the extent of the ion chain at this point is approximately \SI{47}{\micro\meter}. Due to the finite size of the cooling beam, the outer ions are illuminated with a third of the intensity of those in the centre. Measurements were taken with a peak intensity of $I\approx23I_{\mathrm{sat}}$ resulting in the ions fluorescing sufficiently brightly to be counted. 

Across a range of loading times, data is collected for a number of repetitions at the specified heating power. We determine the loading rate by averaging the number of ions counted for a given loading time and linearly fitting the resulting time dependent dataset. Using this method, we measure loading rates from \SI{2.7(8)e-3}{\per\second} up to \SI{24(3)}{\per\second} at heating powers of \SI{37.0(4)}{\milli\watt} and \SI{84.7(8)}{\milli\watt}, respectively. This range of rates is sufficient to repopulate a large array of ions sufficiently quickly enough to enable continuous fault tolerant operation with minimal heat load~\cite{Bruzewicz2016}.

\subsection{Dependence of loading rate on second-stage photo-ionisation laser power}

As the second step of the photo-ionisation process at 375 nm couples to a continuum of unbound states, this process is not saturable. In order to explore the dependence of the loading rate on ionisation laser intensity, the heating beam power was fixed and the ionisation laser power varied. Initially, an ionisation laser intensity of \SI{5.3(4)}{\watt\per\centi\meter\squared} was used. The loading rate was found to be \SI{1.0(2)}{\per\second}. Increasing the ionisation laser intensity to \SI{11.9(6)}{\watt\per\centi\meter\squared} yielded a loading rate of \SI{2.1(2)}{\per\second}. This increase follows the expected linear trend in the ionisation rate with second stage photo-ionisation power consistent with~\cite{Lucas2004}. This shows a clear path to increasing loading rate above the experimentally measured maximum of \SI{24(3)}{\per\second}.

\subsection{Ionisation probability}
To quantify the ionisation probability, we determine the linear relationship between measurements of the number density and ion loading rate. As the trap depth (of order \SI{1}{\electronvolt}) used is significantly larger than the mean kinetic energy of the neutral atoms, we expect the loading rate to be limited predominantly by the ionisation probability rather than laser cooling dynamics.
Using the thermal model described in~\autoref{sec:neutral}, we calculate the number densities of neutral atoms that correspond to the heating powers used in the loading rate measurements. The loading rate~\cite{Lucas2004},
\begin{equation}
    R =  q \times \frac{1}{2}a_{40}n \times  \frac{lw^{2}}{t} \ ,
\end{equation}
can be expressed in terms of the ionisation probability, $q$; the relative abundance of $^{40}\mathrm{Ca}$, $a_{40}$; the number density, $n$; the atomic beam width, $l$, the excitation laser beam waist, $w$; and the time spent in the interaction region, $t$. The factor of $1/2$ accounts for the fraction of $^{40}\mathrm{Ca}$ in the $4 \mathrm{s} 4\mathrm{p} \, {}^{1}\mathrm{P}_{1}$ state at the saturation parameter used in these measurements, where $s \approx 20$.

We perform linear orthogonal distance regression to determine the conversion factor from loading rate to number density. We constrain this analysis to the range of heating powers where measurements for both the loading rate and the number density are available.
For heating powers below \SI{58}{\milli\watt}, the signal-to-noise ratio of neutral density measurements was limiting; for powers above \SI{85}{\milli\watt}, changes in conformation of the trapped ion crystals limited the reliability of loading rate measurements.
The experimentally measured loading rate is shown alongside our inferred number density model in~\hyperref[fig:loading_rate_number_density]{\autoref*{fig:loading_rate_number_density}a}. We indicate the bounds we use to determine the relationship between loading rate and number density with vertical dashed lines. Using this relationship across all data points in \hyperref[fig:loading_rate_number_density]{\autoref*{fig:loading_rate_number_density}a}, we find $q = \num{1.50(5)e-5}$. 
We can extend our model of the neutral density down to the lower heating powers and compare to the loading rate data. In so doing, we find excellent agreement to our thermal model, even at powers as low as \SI{41.4(4)}{\milli\watt}, where a direct measurement of the neutral flux was not feasible.

\section{Conclusions and outlook}\label{sec:conclusions}

The loading rate realised depends on a multitude of experiment-specific parameters, that can be hard to decouple. In our experiments, the trap was sufficiently deep that we can neglect contributions from kinetic energy. The narrow collimation further ensures that the vast majority of atoms that effuse from the oven pass through the trapping region. The high loading rate we demonstrated, however, is still hindered by the fraction of the atomic flux that we can address with the excitation and ionisation lasers. This depends not only on the available laser intensities but also on the interaction volume, which is set by the laser diameters and the spread of the atomic beam.

The ionisation laser intensity has a considerable impact on the achievable loading rates. By extrapolating from our data, we can determine a regime where $q\approx1$, where nearly every excited atom is ionised. This regime could be accessed for ionisation laser intensities of order \SI{e4}{\watt\per\square\centi\meter}, increasing the loading rate by a factor of more than 50,000. Moreover, significant gains could be made by selecting a different ionisation laser wavelength as the photo-ionisation cross section has a wavelength dependence~\cite{Webster2005}. These improvements would culminate in much faster loading rates than those presented.

This style of atomic source allows for a novel mode of operation: by continuously operating at comparatively low heating powers a significant density is still achieved at the trap. The addition of a high-intensity pulsed ionisation laser could therefore provide trappable ions in less than \SI{1}{\milli\second}. This would eliminate any turn-on latency and the design, presented here, significantly reduces the heat load to the experimental apparatus improving the overall thermal stability whilst still allowing for on-demand replacement of lost ions. Moreover, owing to the collimation of the neutral beam, a target atomic density in the trap can be reached at a lower total atomic flux and lower kinetic energy compared to an uncollimated oven, increasing the overall efficiency of the source.

To test the applicability of our source to metals beyond calcium we consider that, for \SI{42}{\milli\watt} of heating laser power, we measured a loading time of approximately \SI{30}{\second}; which compares well with resistively heated sources~\cite{Ballance2018}. Our thermal model predicts that the internal crucible temperature is, for this heating power, approximately \SI{500}{\kelvin}. At this temperature the vapour pressure of calcium inside the oven is of order \SI{e-10}{\milli\bar}~\cite{Alcock1984}. At this temperature the vapour pressure of magnesium, strontium and ytterbium are all higher by at least one order of magnitude~\cite{Alcock1984}, which strongly implies the suitablility for use in ion traps built around these species, even for cryogenic applications. 

If we instead consider \SI{e-10}{\milli\bar} to be a minimum acceptable vapour pressure and allow the heating laser power to increase we can consider the application to a greater range of materials. We find that barium, which requires a crucible temperature of approximately \SI{515}{\kelvin}~\cite{Alcock1984}, also produces suitable densities for loading for only a modest (less than \SI{10}{\milli\watt}) increase in power. We can conclude that, through the thermally isolated design presented, we have produced a source well-suited to the needs of ion-trap researchers using the heavier group-II metals and Yb.

Further, by extrapolating our thermal model to higher laser powers, we predict that less than \SI{1}{\watt} of input laser power is necessary for temperatures of approximately \SI{1000}{\kelvin}. Whilst this is significantly higher in laser power than we use in this work, it is within reach of many experiments. Naturally, our system is unsuited to testing at such high temperatures, as the vapour pressure of calcium would reach such a high level that it would likely damage the structure of the oven. However, more exotic metals used in state-of-the-art ion trapping systems, such as beryllium~\cite{Gaebler2016}, aluminium~\cite{Brewer2019} or lutetium~\cite{Tan2019} would present suitable vapour pressures for loading. Indeed, expanded use of this technique may open up hitherto unprecedented species to being used in ion trap experiments.

In summary, in this work, we have demonstrated an efficient micro-fabricated atom source, that requires no electrical connections. We used this oven for the generation of a neutral beam of $^{40}$Ca atoms, and shown that the beam is of a sufficient peak density to rapidly load ions into a Paul trap, even for modest heating powers. Further, by using a two-photon ionisation technique, we demonstrated the capability to produce atomic fluxes that allow for reliable, continuous re-loading of ion-trap-based quantum processors.

\section*{Data Availability}
The data and analysis that were generated in this work are available at DOI:10.5287/ora-44yqykazk.
\section*{Rights Retention}
This research was funded in whole, or in part, by a Plan S funder. For the purpose of Open Access, the author has applied a CC BY public copyright licence to any Author Accepted Manuscript version arising from this submission.
\section*{Acknowledgments}
This work was supported by: the Engineering and Physical Sciences Research Council (EPSRC) hub in Quantum Computing and Simulation (QCS) [EP/T001062/1]; EPSRC funding under the Horizon Europe Guarantee [EP/Y026438/1] for the European Research Council (ERC) selected and approved ERC Starting Grant: MICRON-QC [101077098]; EPSRC and Science and Technologies Facilities Council Impact accelerator account [EP/X525777/1]. 

LV acknowledges support from Saint Anne's College, Oxford. 
CEJC acknowledges support from the UK National Quantum Computing Centre.
PD acknowledges support from Innovate UK Collaborative Research and Development project LINQED [10100964].
JAB acknowledges support from the EPSRC hub ``Quantum Computing via Integrated and Interconnected Implementations'' (QCI$^3$) [EP/Y024389/1].\\
PD, LJS, JFG are directors of Quantum Fabrix Ltd. (QFX). QFX was not involved in directly funding this work. 
\bibliography{picooven.bib}
\clearpage
\appendix

\section{Steady-state excited population of a calcium atom}\label{app:steady-state}

We consider the manifold consisting of the $^1S_0$ and the three $^1P_1$ states coupled by a laser of frequency $\omega$ and detuning $\delta$. In this discussion, we label with $|g\rangle$ the $|^1 S_0, m_J = 0\rangle$ state and with $|q\rangle$ the $|^1 P_1, m_J = q\rangle$ states for $q\in\{0, -1, +1\}$.

The steady state, if exists, can be found by solving the Lindblad master equation \cite{manzano_short_2020}
\begin{equation}\label{eq:lindblad}
    \dot\rho = -\frac{i}{\hbar}[H, \rho] + A_{\mathrm{SP}} \sum_q (L_q \rho L_q^\dagger - \frac{1}{2}\{L_qL_q^\dagger, \rho\}),
\end{equation}
where the sum is over all possible decay pathways from the $^1P_1$ level and the decay operators are
\begin{equation}
    L_q = |g\rangle\langle q |.
\end{equation}
The Hamiltonian figuring in \eqref{eq:lindblad} that drives the dynamics of the system can be separated into
\begin{equation}
    H = H_0 + H_L
\end{equation}
where the atomic Hamiltonian operator
\begin{equation}
    H_0 = \sum_q \hbar(\omega_0 + q\Delta) |q\rangle\langle q|
\end{equation}
represents the energy of the atom immersed in a magnetic field and $\Delta$ is the Zeeman splitting between the levels in the $^1P_1$ state. The other Hamiltonian operator  $H_L$ describes the laser interaction. To derive an expression for it, we write the electric field associated with the laser field as 
\begin{equation}
    \mathcal{\vec E}(t) = \mathrm{Re}(\vec \epsilon \mathcal{E}_0 e^{-i \omega t}),
\end{equation}
where we have implicitly assumed that the electric field is uniform across the extent of each atom (also known as dipole approximation). The electric dipole energy then reads as
\begin{equation}
   e \hat{\vec r}\cdot \mathcal{\vec E}(t).
\end{equation}

If we write the Hamiltonian operator in the atom eigen-basis and apply the rotating wave approximation, we obtain the expression
\begin{equation}
    H_L(t) = \hbar\sum_q |g\rangle\langle q | \frac{\Omega_q}{2}e^{-i\omega t} + h.c.,
\end{equation}
where the Rabi frequency $\Omega_q$ depends on a combination of laser polarisation and power:
\begin{equation}
    \Omega_q = \frac{e\mathcal{E}_0}{\hbar} \vec \epsilon \cdot \langle g | \hat{\vec r} | q\rangle.
\end{equation}
This system admits a steady state which can be easily found once the time dependence is removed from the Hamiltonian.  This is done by moving into the rotating frame of the laser defined by the unitary operator $U_R=\exp{(-i H_R t/\hbar)}$ which is generated by the Hamiltonian
\begin{equation}
    H_R = -\hbar \omega\sum_q  |q\rangle\langle q|.
\end{equation}
In the interaction picture, the system Hamiltonian becomes
\begin{align}
    H' =&\ U_R^\dagger (H_0 + H_L)U_R - H_R\\
    =& \left( \sum_q |g\rangle\langle q| \frac{\Omega_q}{2} + h.c.  \right) + \hbar \sum_{q}(q\Delta  - \delta)|q\rangle\langle q|,
\end{align}
which is now time-independent. The total population $P_\mathrm{T}'$ across all three sublevels in the $^1P_1$ manifold can be obtained by solving \eqref{eq:lindblad} as a function of laser saturation parameter $s$ and detuning $\delta$ \cite{QuantumToolbox.jl2025}. 
In order to account for the Doppler broadening caused by the $1^\circ$ angle of the neutral excitation beam with the oven axis, we take the ensemble average of $P_T'$ over the speed of the particles. The velocity distribution out of the oven stems from the interaction of the atoms with the walls of the collimator. This is a complicated process to model which is dependent on the surface roughness and finish of the glass surface. For simplicity, we assume elastic collisions to take place, so that the thermal distribution is conserved through the collimator and it is set by the thermal distribution characteristic of an effusion process \cite{landau_statistical}:
\begin{equation}\label{eq:thermal-dist}
    p(v; T) dv = \frac{1}{2}\left(\frac{m_{\mathrm{Ca}_{40}}}{k_B T}\right)^2 e^{-\frac{m_{\mathrm{Ca}_{40}}v^2}{2 k_BT}} v^3,
\end{equation}
where $p(v; T)$ is normalised over the interval $v\in[0,\infty)$. The expression for the total excited population averaged across the thermal ensemble of atoms is given by
\begin{equation}\label{eq:excited-pop}
    P_\mathrm{T}(s; \delta) = \int p(v; T)dv P_T'(s, \delta_a(v) + \delta),
\end{equation}
where $\delta_a(v)=\omega_0 v/c\cdot \sin(1^\circ)$ is the Doppler shift experienced by an atom travelling at velocity $v$. The laser detuning $\delta$ was experimentally set to maximise the fluoresce signal while probing the atoms with $s\approx 1$, while the oven operated around an approximate temperature of $T=550\ \mathrm{K}$. Following from \eqref{eq:excited-pop}, this corresponds to a detuning of $\delta_{\mathrm{max}} = -(2\pi)23.6\ \mathrm{MHz}$ from resonance.  The final expression for $P_\mathrm{T}$ as a function of the saturation parameter only, is then given by:
\begin{equation}
    P_\mathrm{T}(s) = P_\mathrm{T}(s; \delta_{\mathrm{max}}),
\end{equation}
which is used throughout the analysis in the main text.

\section{Obtaining the internal flux from the measured density}\label{app:flux}

\begin{figure}
    \centering
    \includegraphics[width=\linewidth]{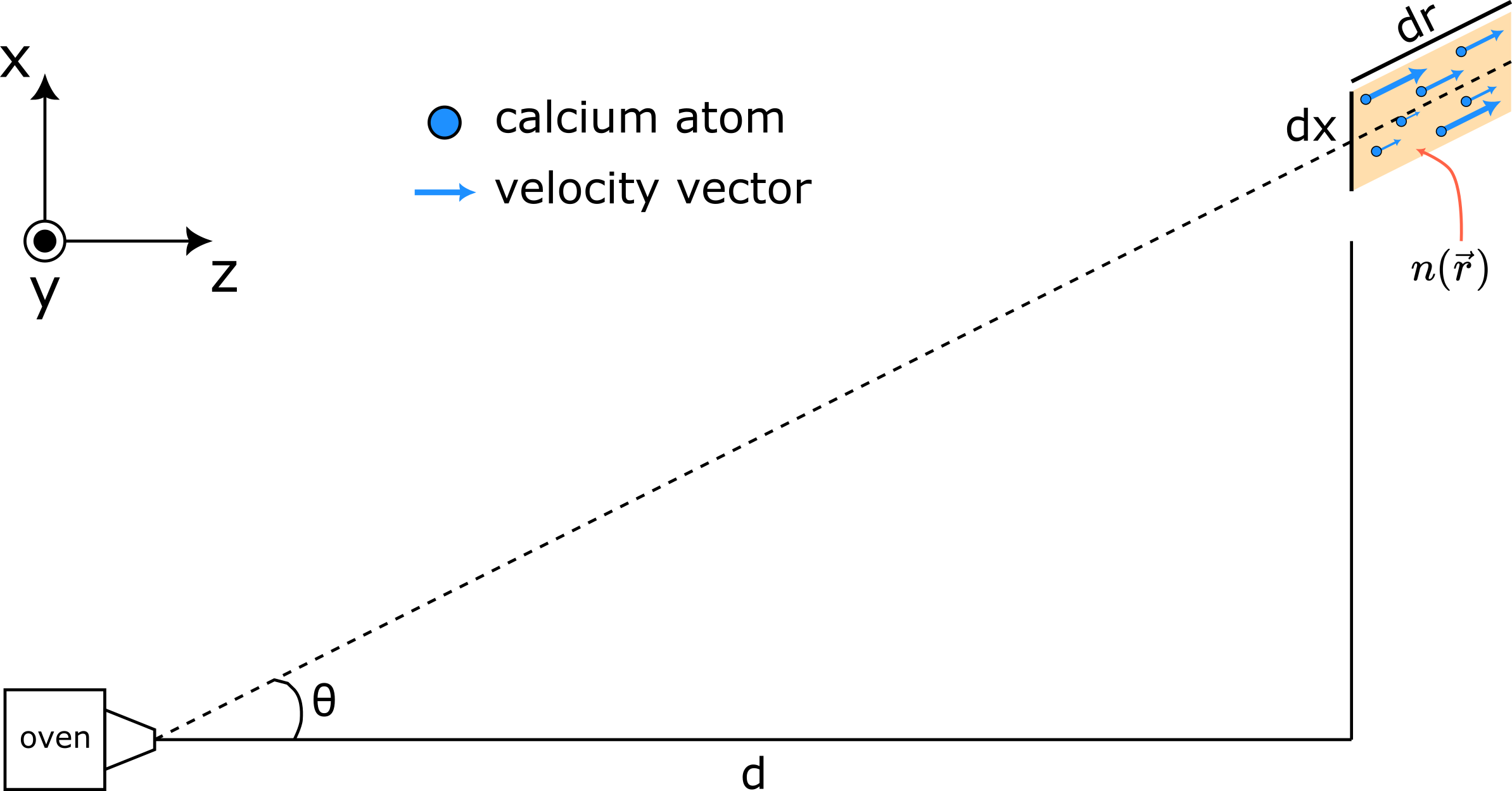}
    \caption{Diagrammatic representation of the atomic flux coming from the oven. The flux out of the oven can be inferred from the known density and the velocity distribution at a plane set at a distance $d$ from the crucible. Given that $d\gg50\ \mathrm{\upmu m}$ (collimator diameter), the oven is modelled as a point source.}
    \label{fig:flux-formula}
\end{figure}

In order to obtain the temperature inside the crucible, we need to obtain an expression for the atomic flux leaving the oven. We can convert the density of $^{40}\mathrm{Ca}$ into an atomic flux using an appropriate thermal velocity distribution for the atoms. On a plane at a distance $d$ from the oven, we have approximated the density of emitters to be
\begin{equation}
    n(\vec r) = n_{\mathrm{peak}} e^{-\frac{2(x^2+y^2)}{w_{\mathrm{a}}^2}}.
\end{equation}
We are interested in finding the flux through this plane and in order to do so, we consider the infinitesimal volume element depicted in \autoref{fig:flux-formula} and the number of atoms with velocity $v\in [v, v+dv]$ contained within, which is given by the formula
\begin{equation}
    \mathrm{d}x\ \mathrm{d}y\ \mathrm{d}r\ \cos(\theta) n(\vec r)p(v; T) \mathrm{d}v,
\end{equation}
where $p(v; T)$ is the velocity probability distribution. If we divide this expression by the time $\mathrm{d}t$ it takes for the atom to traverse $\mathrm{d}r$, we get the partial contribution of this small surface element to the flux
\begin{equation}
    \mathrm{d}x\ \mathrm{d}y\ \cos(\theta) n(\vec r) p(\vec v; T) v\ \mathrm{d}v.
\end{equation}
Integrating this over the velocities $v$, and the spatial coordinates $x$ and $y$ gives the total flux through the plane. By noting that the average velocity is just $\bar v = \int_0^\infty p(\vec v; T) v\ \mathrm{d}v$, $\cos(\theta)= d/\sqrt{x^2+y^2+d^2}$, and by correcting the density by the natural abundance $a_{40}$, we can write the flux out of the oven as:
\begin{align}
    \Phi_{\mathrm{out}} &= \bar v \int_{-\infty}^{\infty}\int_{-\infty}^{\infty}\mathrm{d}x \mathrm{d}y \frac{d}{\sqrt{x^2+y^2+d^2}} \frac{n_{\mathrm{peak}}}{a_{40}}e^{-\frac{2(x^2 + y^2)}{ w_{\mathrm{a}}^2}}\\
    &=2\pi \bar v \frac{n_{\mathrm{peak}}}{a_{40}}\int_{0}^{\infty}d\rho \frac{\rho d}{\sqrt{\rho^2 + d^2}}e^{-\frac{2\rho^2}{w_{\mathrm{a}}^2}}\\
    &=\frac{1}{2}\pi \bar v n_{\mathrm{peak}} w_{\mathrm{a}}^2 \mathcal{I}/a_{40},\label{eq:flux-out}
\end{align}
where in the second line we have performed the change of variable from $(x,y)$ to polar coordinates $(\rho, \phi)$, with the integration over $\phi$ giving the factor of $2\pi$, and in the third line we have defined the dimensionless integral $\mathcal{I}$ through the additional change of variable $\alpha = 2r/w_{\mathrm{a}}$ as
\begin{equation}
    \mathcal{I} = \int_0^\infty d\alpha \frac{2\alpha d/w_{\mathrm{a}}}{\sqrt{\alpha^2+(2d/w_{\mathrm{a}})^2}}e^{-\alpha^2/2}.
\end{equation}\\[1mm]
Due to the atomic beam being highly collimated, this integral evaluates to $\mathcal{I}=0.99982(4)$  for our experimental parameters and can therefore be neglected. The expression for the output flux is then greatly simplified and it evaluates to the product of the average output velocity $\bar v$ and the integral of the density $n(\vec r)/a_{40}$ over the $(x,y)$ plane, which evaluates to $\frac{1}{2}\pi w_{\mathrm{a}}^2 n_{\mathrm{peak}}/a_{40}$. The average velocity can be obtained from the thermal distribution \eqref{eq:thermal-dist}
\begin{equation}
    \bar v = \sqrt{\frac{9 \pi k_B T}{8m_{\mathrm{ca}}}}.
\end{equation}
The flux incident into the collimator on the crucible side is \cite{landau_statistical}
\begin{equation}\label{eq:flux-in}
    \Phi_{\mathrm{in}} = \frac{P(T) A_{\mathrm{coll}}}{\sqrt{2\pi m_{\mathrm{ca}}k_{B}T}},
\end{equation}
where $A_{\mathrm{coll}}=\pi (25\ \mathrm{\upmu m})^2$ is the surface are of the collimator aperture. Given that the probability of a particle being transmitted through the collimator and emerging from the other side is $p_T$, we can equate the input flux and the output flux as $\Phi_{\mathrm{out}} = p_t \Phi_{\mathrm{in}}$. By substituting \eqref{eq:flux-in} and \eqref{eq:flux-out}, we can obtain an expression that links the temperature inside the crucible to the measured peak density value:
\begin{equation}
    \frac{P(T)}{k_B T}=\frac{3\pi^2}{4p_t}\frac{w_{\mathrm{a}}^2}{A_{\mathrm{coll}}} \frac{n_{\mathrm{peak}}}{a_{40}},
\end{equation}
which can be solved numerically for the crucible temperature $T$, where $P(T)$, the vapour pressure of calcium, is well approximated by 
\begin{equation}
    \log_{10}(P/1\ \mathrm{atm)} = A + \frac{B}{T} + C\log_{10}(T/1\ \mathrm{K)} + DT/1000
\end{equation}
with the coefficients given in \cite{Alcock1984}.

\end{document}